\documentclass[3p,times,procedia]{elsarticle}
\usepackage{nupha_ecrc}

\usepackage{bm}

\newcommand{\state}[4]{{^#1\hspace{-0.6mm}#2_{#3}^{[#4]}}}

\volume{00}

\firstpage{1}

\journalname{Nuclear Physics A}

\runauth{}

\jid{nupha}

\jnltitlelogo{Nuclear Physics A}

\usepackage{amssymb}

\usepackage[figuresright]{rotating}

\begin{document}

\begin{frontmatter}



\dochead{XXVIIth International Conference on Ultrarelativistic Nucleus-Nucleus Collisions\\ (Quark Matter 2018)}

\title{Unified framework for heavy flavor and quarkonium production in high multiplicity p+p and p+A collisions at RHIC and LHC}


\author[label1,label2,label3]{Yan-Qing Ma}
\author[label4]{Prithwish Tribedy}
\author[label4]{Raju Venugopalan}
\author[label5]{Kazuhiro Watanabe}

\address[label1]{School of Physics and State Key Laboratory of Nuclear Physics and Technology, Peking University, Beijing 100871, China.}
\address[label2]{Center for High Energy Physics, Peking University, Beijing 100871, China.}
\address[label3]{Collaborative Innovation Center of Quantum Matter, Beijing 100871, China.}
\address[label4]{Physics Department, Brookhaven National Laboratory, Upton, New York 11973, USA.}
\address[label5]{Theory Center, Thomas Jefferson National Accelerator Facility, Newport News, Virginia 23606, USA.}

\begin{abstract}
We discuss the production of $D$-mesons and $J/\psi$ in high multiplicity proton-proton and proton-nucleus collisions within  the Color-Glass-Condensate (CGC) framework. We demonstrate that the modification of the LHC data on $D$ and $J/\psi$ yields in high multiplicity events relative to minimum bias events arise from a significant enhancement of the 
gluon saturation scales of the corresponding rare parton configurations in the colliding protons and nuclei. For a given event multiplicity, we predict these relative yields to be energy independent from $\sqrt{s}=200$ GeV at RHIC to the highest LHC energies. 
\end{abstract}


\end{frontmatter}



\section{Introduction}\label{Introduction}

Heavy flavor ($D$) and quarkonium ($J/\psi$) production in minimum bias proton-proton($p+p$) and proton-nucleus ($p+A$) collisions at RHIC and LHC have been studied extensively in the Color-Glass-Condensate (CGC) effective field theory over the last several years~\cite{Ma:2014mri,Ma:2015sia,Ma:2017rsu,Watanabe:2015yca,Fujii:2013yja,Fujii:2015lld,Fujii:2017rqa}. These works demonstrated that quantitative descriptions of the transverse momentum and rapidity distributions could be achieved in these collisions. The study of rare high multiplicity events in $p+p$ and $p+A$ collisions at RHIC and the LHC have opened up a new window in the dynamics of atypical parton configurations in QCD. A striking result in such events are the ridge-like structures of azimuthally collimated and long range in rapidity hadron correlations leading to an interesting debate on the relative importance of initial or final state effects. Systematic studies of heavy quarks-- a so-called ``event engineering" can play an important role in clarifying the underlying mechanisms of high multiplicity events. In this proceedings, we will present a brief perspective on the comparison of  currently available experimental data on high multiplicity $D$ meson and $J/\psi$ production in $p+p$ and $p+A$ collisions~\cite{Adam:2015ota,Adam:2016mkz,Abelev:2012rz,Weber:2017hhm,Khatun:2017yic} with our recent results~\cite{Ma:2018bax} obtained within the CGC EFT.

\section{CGC Framework}\label{Framework}

\begin{figure*}
\centering
\includegraphics[width=0.45\linewidth]{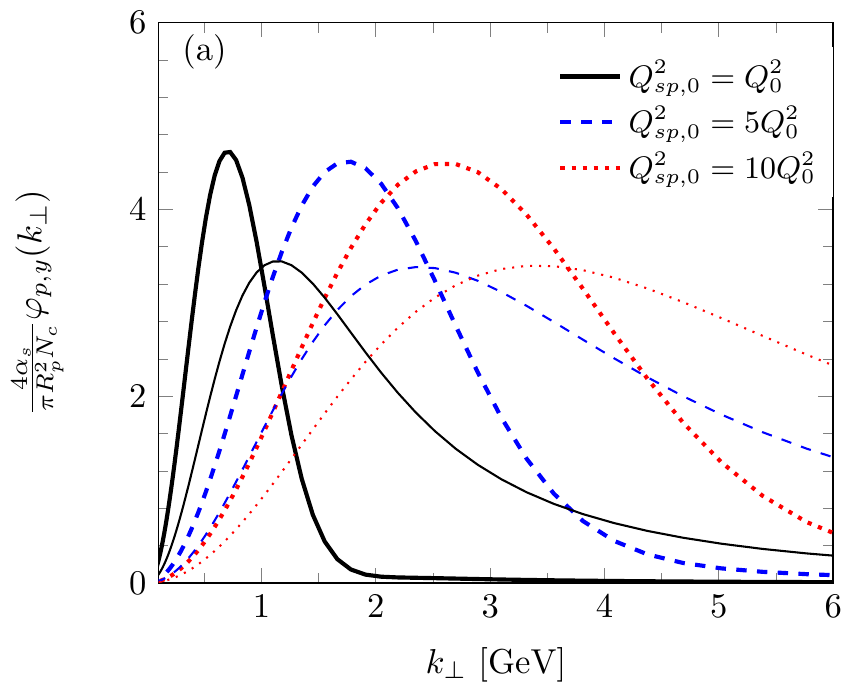}
\includegraphics[width=0.45\linewidth]{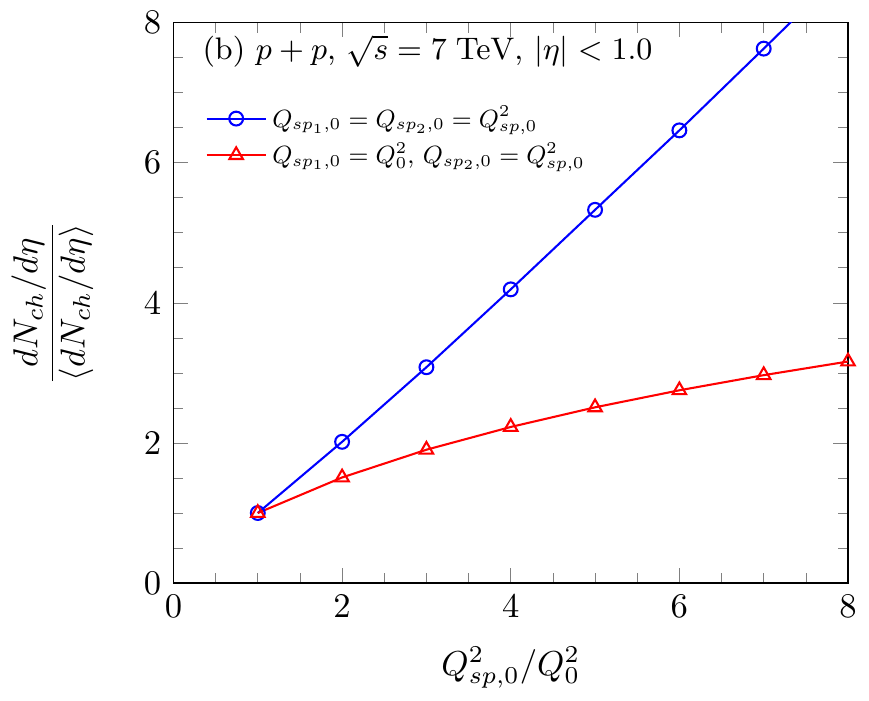}
\caption{
(a): $k_\perp$ distribution of $\varphi_p$ at $x=0.01$ (thick) and $x=10^{-4}$ (thin) by varying the initial saturation scale $Q_{sp,0}^2$. (b): $Q_{sp,0}^2$ vs normalized charged hadron multiplicity $dN_{ch}/\langle dN_{ch}\rangle$ in $p+p$ collisions at the LHC. The circular points (in blue) are results for the normalized charged particle multiplicity obtained by equating the saturation scales for the projectile and target: $Q_{sp_1,0}=Q_{sp_2,0}$. The triangular points (in red) are obtained by varying $Q_{sp_2,0}^2$ while $Q_{sp_1,0}^2$ is fixed as $Q_0^2$.}
\label{fig:Qs-dependence}
\end{figure*}

We shall focus on the charged hadron multiplicity $N_{ch}$ dependence of $D$ and $J/\psi$ production in this paper. In our calculations, the charged hadron production in $p+p(A)$ collisions is computed by convoluting inclusive gluon production cross-section~\cite{Kovchegov:2001sc}
\begin{eqnarray}
\frac{d\sigma_{g}}{d^2\bm{p}_{g\perp} dy}=\frac{\alpha_s}{(2\pi)^3\pi^3 C_F}\frac{1}{{p}_{g\perp}^2}\int d^2\bm{k}_\perp \varphi_{p,y_p}(\bm{k}_\perp)\varphi_{A,Y}(\bm{p}_{g\perp}-\bm{k}_\perp)\,,
\label{eq:gluon-kt}
\end{eqnarray}
with the KKP fragmentation function (FF)~\cite{Kniehl:2000fe}. Here $y_p=\ln1/x_1$ and $Y=\ln1/x_2$ with $x_{1,2}$ being incoming gluon's longitudinal momentum fraction. Information on impact parameter is encoded in the unintegrated gluon distribution function (UGDF) $\varphi$ through the saturation scales of the proton and nucleus. 

The differential cross-section for $c\bar c$ production in $p+p(A)$ collisions in the large $N_c$ limit can be expressed as~\cite{Blaizot:2004wv}
\begin{eqnarray}
\frac{d \sigma_{c \bar{c}}}{d^2\bm{p}_{c\perp} d^2\bm{q}_{\bar c\perp} dy_c dy_{\bar c}}
=
\frac{\alpha_s N_c^2 \pi R_{A}^2}{2(2\pi)^{10} (N_c^2-1)}
\int d^2\bm{k}_{2\perp}d^2\bm{k}_\perp
\frac{\varphi_{p,y_p}(\bm{k}_{1\perp})}
{k_{1\perp}^2}
\mathcal{N}_{Y}(\bm{k}_\perp)\mathcal{N}_{Y}(\bm{k}_{2\perp}-\bm{k}_\perp)\, \Xi,
\label{eq:xsection-kt-factorization-LN}
\end{eqnarray}
where $\Xi$ is hard part and $\mathcal{N}_{Y}$ is the fundamental dipole amplitude. Single $D$ meson production is straightforwardly calculated by using heavy quark FF which is set to be the QCD evolved BCFY FF~\cite{Cacciari:2012ny}. In computing the $J/\psi$ cross-section, we will employ the CGC + NRQCD formalism~\cite{Kang:2013hta}, wherein the  hard matrix elements differ from $\Xi$, as well as the CGC+ ICEM model, whereby the latter is the improved color evaporation model~\cite{Ma:2016exq}.

The rapidity dependence of the cross-sections is computed by solving the running coupling BK equation~\cite{Balitsky:2006wa} for $\mathcal{N}_{Y}$. All input parameters in the initial dipole amplitude at $x_0=0.01$ are determined by DIS global fits to HERA data. In our numerical computations, we take the initial saturation scale $Q_{sp,0}^2=0.168\,\textrm{GeV}^2$ for the proton and $Q_{sA,0}^2=2\,Q_{sp,0}^2$ for nuclei as discussed in~\cite{Ma:2015sia,Ma:2017rsu}.  

High values of $N_{ch}$ are achieved by allowing the initial saturation scales of both the projectile and target to fluctuate to large values, corresponding to a fluctuation of large $x$ parton configurations in these rare events.  Figure~\ref{fig:Qs-dependence} (a) shows the $\varphi_p$ distribution in such events at both the initial saturation scale of $x=0.01$ and after evolution to $x=10^{-4}$. We observe that QCD evolution in the rare events leads to very broad distributions peaked at values of transverse momenta that are $\sim 3$ GeV. If the saturation scales of the projectile and target are similar, $\varphi_{p_1}$ and $\varphi_{p_2}$ will have a large overlapping weight in the integrand of Eq.~\ref{eq:gluon-kt} leading to a  rapid increase of $N_{ch}$ as shown in Fig.~\ref{fig:Qs-dependence} (b).

\section{Results}\label{Results}

\begin{figure*}[t]
\centering
\includegraphics[width=0.45\linewidth]{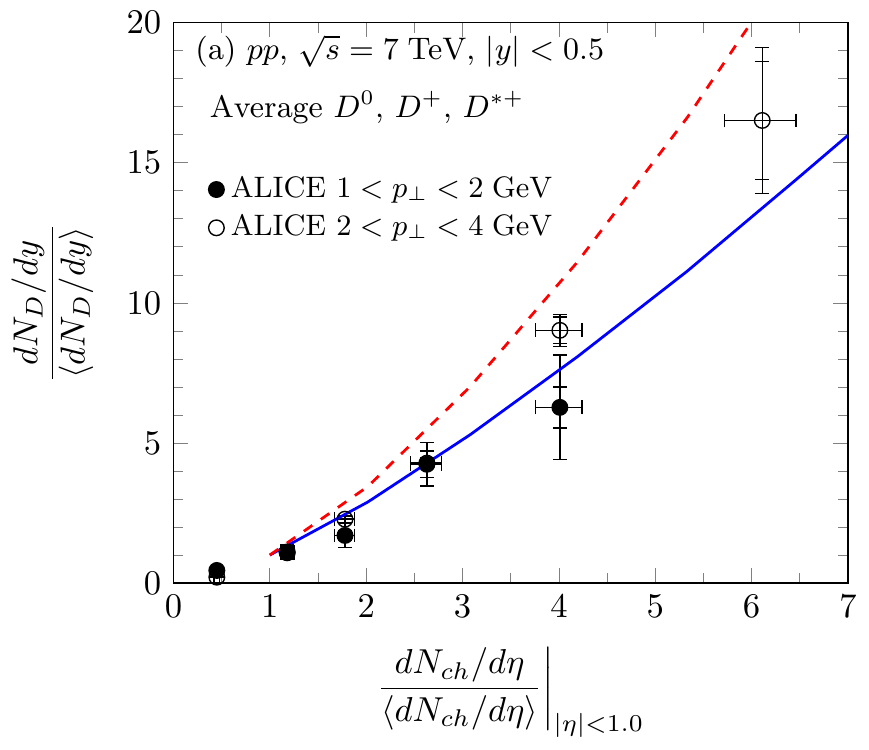}
\includegraphics[width=0.45\linewidth]{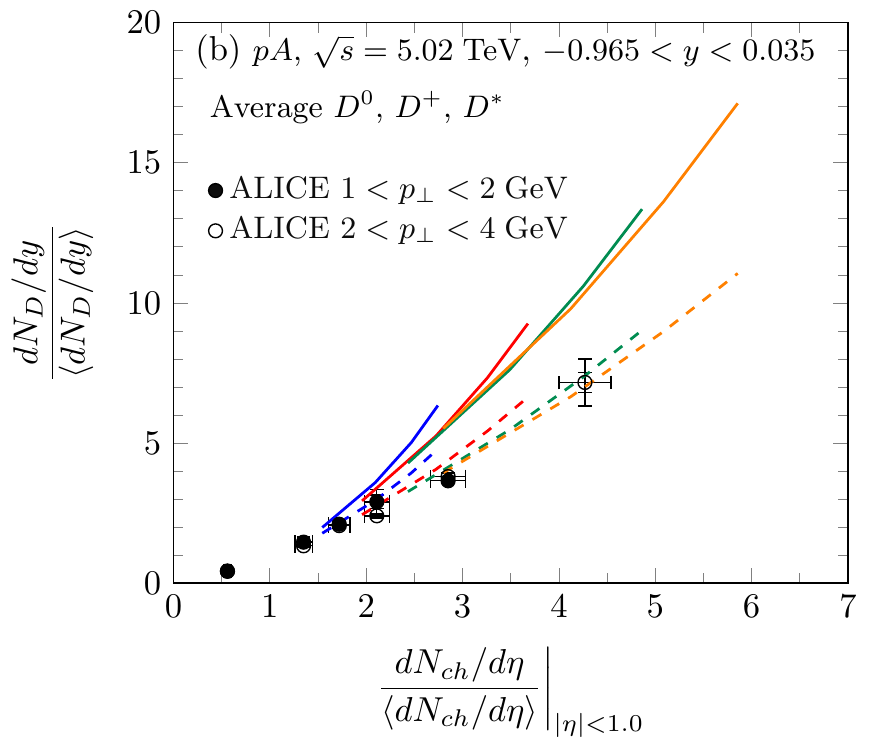}
\caption{
(a): Relative yields of average $D$ ($D^0$, $D^+$, $D^{\ast+}$) as a function of relative multiplicity in $p+p$ collisions at the LHC. The solid (dashed) line is the result at $1<p_\perp<2\,\textrm{GeV}$ ($2<p_\perp<4\,\textrm{GeV}$). Data are from Ref.~\cite{Adam:2015ota}.
(b): Results in $p+A$ collisions. The solid (dashed) lines are the results at $1<p_\perp<2\,\textrm{GeV}$ ($2<p_\perp<4\,\textrm{GeV}$). The blue, red, green and orange curves all show model results for variations in the range $Q_{sp,0}^2 = 1-4 Q_0^2$ for $Q_{sA,0}^2=4,6,9,12 Q_0^2$ respectively. Data are from Ref.~\cite{Adam:2016mkz}.}
\label{fig:D-Nch}
\end{figure*}
\begin{figure*}[h]
\centering
\includegraphics[width=0.325\linewidth]{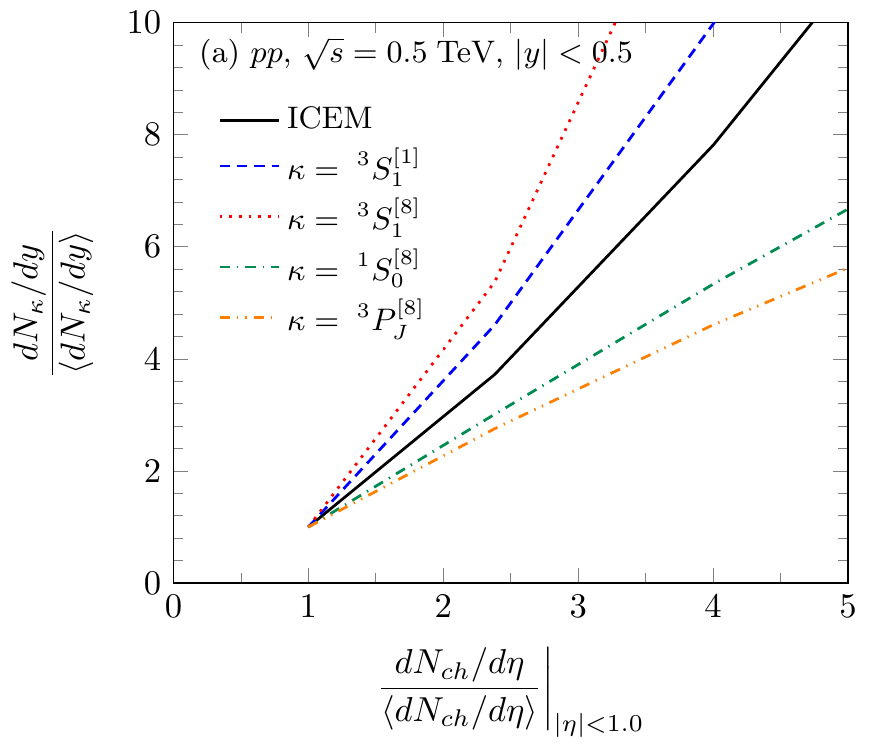}
\includegraphics[width=0.325\linewidth]{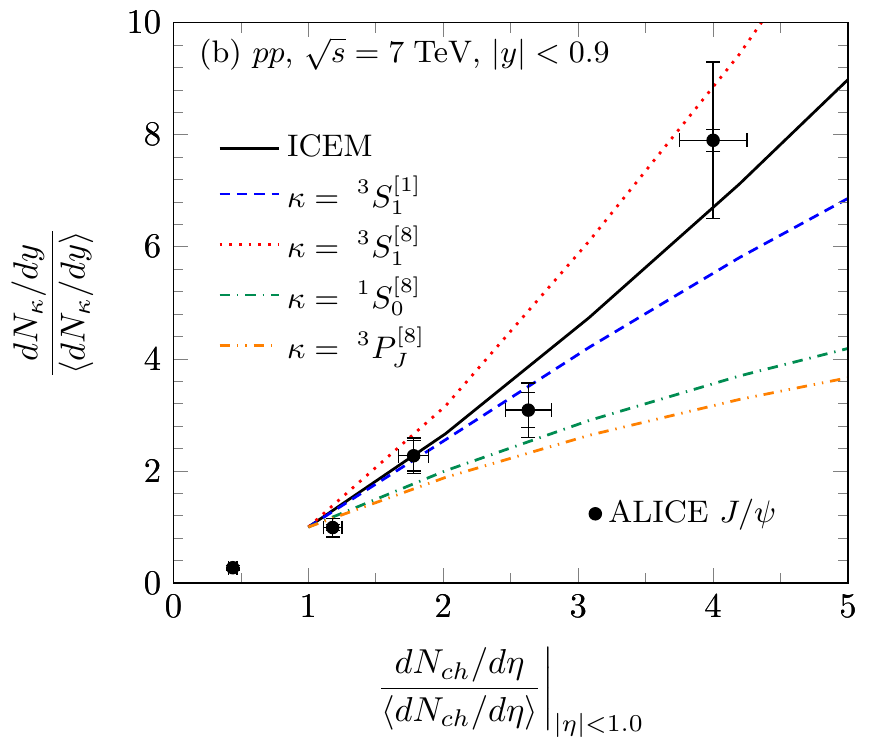}
\includegraphics[width=0.325\linewidth]{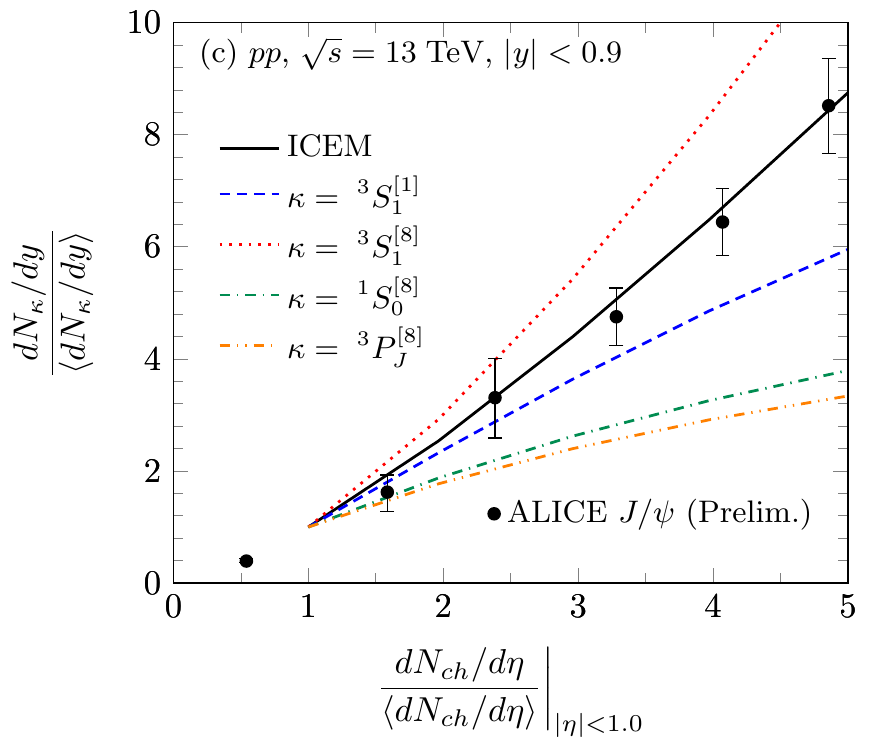}
\caption{
$N_{ch}$ dependence of $J/\psi$ production in $p+p$ collisions in the CGC+ICEM (solid curve) and the CGC+NRQCD (other curves) at RHIC (a) and the LHC (b)(c). Data points are from Refs.~\cite{Abelev:2012rz,Weber:2017hhm,Khatun:2017yic}.
}
\label{fig:Jpsi-Nch-pp+pA}
\end{figure*}

In Fig.~\ref{fig:D-Nch}, we present our results on the $N_{ch}$ dependence of $D$ production in both $p+p$ and $p+A$ collisions at the LHC. The CGC EFT shows a nice agreement with the data within the systematic uncertainties with respect to the FFs and the initial saturation scales. These are however quite small for the double ratios presented. A similar trend of the  $p_\perp$ dependence of the double ratios is seen in both $p+p$ and $p+A$ collisions~\cite{Ma:2018bax}.

The $N_{ch}$ dependence of $J/\psi$ production in $p+p$ collisions is shown in Fig.~\ref{fig:Jpsi-Nch-pp+pA}. The relative contribution of different intermediate channels $\kappa=\state{{2S+1}}{L}{J}{c}$ is examined in the CGC+NRQCD framework. We find that the $^3S_1^{[8]}$ intermediate state has a larger relative weight than the other channels for all $dN_{\rm ch}/d\eta$. This suggests that $J/\psi$ production with high event activity is dominated by the $^3S_1^{[8]}$ channel but has only small contributions from $^1S_0^{[8]}$ and $^3P_J^{[8]}$ channels. This result is  consistent with the universality requirement for LDMEs employed in hadron-hadron collisions to be consistent with those extracted from BELLE $e^+e^-$ data~\cite{Zhang:2009ym}. In Fig.~\ref{fig:Jpsi-Nch-pp+pA}, we also put the ICEM curves which have trends similar to the $^3S_1^{[8]}$ channel. Comparisons of the CGC+ICEM computations to available data at the LHC indicate that the event activity dependence of the ratios for $J/\psi$ production is insensitive to $\sqrt{s}$. We therefore predict that results for the relative yields as a function of event activity will be identical (within systematic uncertainties) as those from the LHC. However, the relative weight of different $\kappa$ can be changed at RHIC energy. We have also checked that $J/\psi$ production in $p+A$ collisions in the CGC+ICEM provides a nice agreement with data on the ratio~\cite{Ma:2018bax}. 

It will be interesting in future to address event activity dependence of $\psi(2S)$ and $\Upsilon$ in $p+p$ and $p+A$ collisions. However this will require including additional effects such as comover partons and soft gluons emission (Sudakov factor)~\cite{Ma:2017rsu,Watanabe:2015yca} that become increasingly important with increasing heavy quark masses.

\section*{Acknowledgements}
K.W. is supported by Jefferson Science Associates, LLC under  U.S. DOE Contract No.~DE-AC05-06OR23177. P.T. and R.V. are supported by the U.S. Department of Energy Office of Science, Office of Nuclear Physics, under contracts No.~\textrm{DE-SC0012704}.





\bibliographystyle{elsarticle-num}
\bibliography{BibTexData}

\begin{thebibliography}{10}
\expandafter\ifx\csname url\endcsname\relax
  \def\url#1{\texttt{#1}}\fi
\expandafter\ifx\csname urlprefix\endcsname\relax\def\urlprefix{URL }\fi
\expandafter\ifx\csname href\endcsname\relax
  \def\href#1#2{#2} \def\path#1{#1}\fi

\bibitem{Ma:2014mri}
Y.-Q. Ma, R.~Venugopalan, {Comprehensive Description of $J/\psi$ Production in
  Proton-Proton Collisions at Collider Energies}, Phys.Rev.Lett. 113~(19)
  (2014) 192301.
\newblock \href {http://arxiv.org/abs/1408.4075} {\path{arXiv:1408.4075}},
  \href {http://dx.doi.org/10.1103/PhysRevLett.113.192301}
  {\path{doi:10.1103/PhysRevLett.113.192301}}.

\bibitem{Ma:2015sia}
Y.-Q. Ma, R.~Venugopalan, H.-F. Zhang, {$J/\psi$ production and suppression in
  high energy proton-nucleus collisions}, Phys. Rev. D92 (2015) 071901.
\newblock \href {http://arxiv.org/abs/1503.07772} {\path{arXiv:1503.07772}},
  \href {http://dx.doi.org/10.1103/PhysRevD.92.071901}
  {\path{doi:10.1103/PhysRevD.92.071901}}.

\bibitem{Ma:2017rsu}
Y.-Q. Ma, R.~Venugopalan, K.~Watanabe, H.-F. Zhang, {$\psi(2S)$ versus $J/\psi$
  suppression in proton-nucleus collisions from factorization violating soft
  color exchanges}, Phys. Rev. C97~(1) (2018) 014909.
\newblock \href {http://arxiv.org/abs/1707.07266} {\path{arXiv:1707.07266}},
  \href {http://dx.doi.org/10.1103/PhysRevC.97.014909}
  {\path{doi:10.1103/PhysRevC.97.014909}}.

\bibitem{Watanabe:2015yca}
K.~Watanabe, B.-W. Xiao, {Forward Heavy Quarkonium Productions at the LHC},
  Phys. Rev. D92~(11) (2015) 111502.
\newblock \href {http://arxiv.org/abs/1507.06564} {\path{arXiv:1507.06564}},
  \href {http://dx.doi.org/10.1103/PhysRevD.92.111502}
  {\path{doi:10.1103/PhysRevD.92.111502}}.

\bibitem{Fujii:2013yja}
H.~Fujii, K.~Watanabe, {Heavy quark pair production in high energy pA
  collisions: Open heavy flavors}, Nucl. Phys. A920 (2013) 78--93.
\newblock \href {http://arxiv.org/abs/1308.1258} {\path{arXiv:1308.1258}},
  \href {http://dx.doi.org/10.1016/j.nuclphysa.2013.10.006}
  {\path{doi:10.1016/j.nuclphysa.2013.10.006}}.

\bibitem{Fujii:2015lld}
H.~Fujii, K.~Watanabe, {Leptons from heavy-quark semileptonic decay in p A
  collisions within the CGC framework}, Nucl. Phys. A951 (2016) 45--59.
\newblock \href {http://arxiv.org/abs/1511.07698} {\path{arXiv:1511.07698}},
  \href {http://dx.doi.org/10.1016/j.nuclphysa.2016.03.045}
  {\path{doi:10.1016/j.nuclphysa.2016.03.045}}.

\bibitem{Fujii:2017rqa}
H.~Fujii, K.~Watanabe, {Nuclear modification of forward $D$ production in pPb
  collisions at the LHC}\href {http://arxiv.org/abs/1706.06728}
  {\path{arXiv:1706.06728}}.

\bibitem{Adam:2015ota}
J.~Adam, et~al., {Measurement of charm and beauty production at central
  rapidity versus charged-particle multiplicity in proton-proton collisions at
  $ \sqrt{s}=7 $ TeV}, JHEP 09 (2015) 148.
\newblock \href {http://arxiv.org/abs/1505.00664} {\path{arXiv:1505.00664}},
  \href {http://dx.doi.org/10.1007/JHEP09(2015)148}
  {\path{doi:10.1007/JHEP09(2015)148}}.

\bibitem{Adam:2016mkz}
J.~Adam, et~al., {Measurement of D-meson production versus multiplicity in p-Pb
  collisions at $ \sqrt{{\mathrm{s}}_{\mathrm{NN}}}=5.02 $ TeV}, JHEP 08 (2016)
  078.
\newblock \href {http://arxiv.org/abs/1602.07240} {\path{arXiv:1602.07240}},
  \href {http://dx.doi.org/10.1007/JHEP08(2016)078}
  {\path{doi:10.1007/JHEP08(2016)078}}.

\bibitem{Abelev:2012rz}
B.~Abelev, et~al., {$J/\psi$ Production as a Function of Charged Particle
  Multiplicity in $pp$ Collisions at $\sqrt{s} = 7$ TeV}, Phys. Lett. B712
  (2012) 165--175.
\newblock \href {http://arxiv.org/abs/1202.2816} {\path{arXiv:1202.2816}},
  \href {http://dx.doi.org/10.1016/j.physletb.2012.04.052}
  {\path{doi:10.1016/j.physletb.2012.04.052}}.

\bibitem{Weber:2017hhm}
S.~G. Weber, {Measurement of $J/\psi$ production as a function of event
  multiplicity in pp collisions at $\sqrt{s} = 13\,\mathrm{TeV}$ with ALICE},
  Nucl. Phys. A967 (2017) 333--336.
\newblock \href {http://arxiv.org/abs/1704.04735} {\path{arXiv:1704.04735}},
  \href {http://dx.doi.org/10.1016/j.nuclphysa.2017.06.054}
  {\path{doi:10.1016/j.nuclphysa.2017.06.054}}.

\bibitem{Khatun:2017yic}
A.~Khatun, {Measurement of J/$\psi$ production as a function of multiplicity in
  pp and p-Pb collisions with ALICE}.
\newblock \href {http://arxiv.org/abs/1711.09865} {\path{arXiv:1711.09865}}.

\bibitem{Ma:2018bax}
Y.-Q. Ma, P.~Tribedy, R.~Venugopalan, K.~Watanabe, {Event engineering heavy
  flavor production and hadronization in high multiplicity hadron-hadron
  collisions}\href {http://arxiv.org/abs/1803.11093} {\path{arXiv:1803.11093}}.

\bibitem{Kovchegov:2001sc}
Y.~V. Kovchegov, K.~Tuchin, {Inclusive gluon production in DIS at high parton
  density}, Phys. Rev. D65 (2002) 074026.
\newblock \href {http://arxiv.org/abs/hep-ph/0111362}
  {\path{arXiv:hep-ph/0111362}}, \href
  {http://dx.doi.org/10.1103/PhysRevD.65.074026}
  {\path{doi:10.1103/PhysRevD.65.074026}}.

\bibitem{Kniehl:2000fe}
B.~A. Kniehl, G.~Kramer, B.~Potter, {Fragmentation functions for pions, kaons,
  and protons at next-to-leading order}, Nucl. Phys. B582 (2000) 514--536.
\newblock \href {http://arxiv.org/abs/hep-ph/0010289}
  {\path{arXiv:hep-ph/0010289}}, \href
  {http://dx.doi.org/10.1016/S0550-3213(00)00303-5}
  {\path{doi:10.1016/S0550-3213(00)00303-5}}.

\bibitem{Blaizot:2004wv}
J.~P. Blaizot, F.~Gelis, R.~Venugopalan, {High-energy pA collisions in the
  color glass condensate approach. 2. Quark production}, Nucl.Phys. A743 (2004)
  57--91.
\newblock \href {http://arxiv.org/abs/hep-ph/0402257}
  {\path{arXiv:hep-ph/0402257}}, \href
  {http://dx.doi.org/10.1016/j.nuclphysa.2004.07.006}
  {\path{doi:10.1016/j.nuclphysa.2004.07.006}}.

\bibitem{Cacciari:2012ny}
M.~Cacciari, S.~Frixione, N.~Houdeau, M.~L. Mangano, P.~Nason, G.~Ridolfi,
  {Theoretical predictions for charm and bottom production at the LHC}, JHEP 10
  (2012) 137.
\newblock \href {http://arxiv.org/abs/1205.6344} {\path{arXiv:1205.6344}},
  \href {http://dx.doi.org/10.1007/JHEP10(2012)137}
  {\path{doi:10.1007/JHEP10(2012)137}}.

\bibitem{Kneesch:2007ey}
T.~Kneesch, B.~A. Kniehl, G.~Kramer, I.~Schienbein, {Charmed-meson
  fragmentation functions with finite-mass corrections}, Nucl. Phys. B799
  (2008) 34--59.
\newblock \href {http://arxiv.org/abs/0712.0481} {\path{arXiv:0712.0481}},
  \href {http://dx.doi.org/10.1016/j.nuclphysb.2008.02.015}
  {\path{doi:10.1016/j.nuclphysb.2008.02.015}}.

\bibitem{Kang:2013hta}
Z.-B. Kang, Y.-Q. Ma, R.~Venugopalan, {Quarkonium production in high energy
  proton-nucleus collisions: CGC meets NRQCD}, JHEP 1401 (2014) 056.
\newblock \href {http://arxiv.org/abs/1309.7337} {\path{arXiv:1309.7337}},
  \href {http://dx.doi.org/10.1007/JHEP01(2014)056}
  {\path{doi:10.1007/JHEP01(2014)056}}.

\bibitem{Ma:2016exq}
Y.-Q. Ma, R.~Vogt, {Quarkonium Production in an Improved Color Evaporation
  Model}, Phys. Rev. D94~(11) (2016) 114029.
\newblock \href {http://arxiv.org/abs/1609.06042} {\path{arXiv:1609.06042}},
  \href {http://dx.doi.org/10.1103/PhysRevD.94.114029}
  {\path{doi:10.1103/PhysRevD.94.114029}}.

\bibitem{Balitsky:2006wa}
I.~Balitsky, {Quark contribution to the small-x evolution of color dipole},
  Phys. Rev. D75 (2007) 014001.
\newblock \href {http://arxiv.org/abs/hep-ph/0609105}
  {\path{arXiv:hep-ph/0609105}}, \href
  {http://dx.doi.org/10.1103/PhysRevD.75.014001}
  {\path{doi:10.1103/PhysRevD.75.014001}}.

\bibitem{Zhang:2009ym}
Y.-J. Zhang, Y.-Q. Ma, K.~Wang, K.-T. Chao, {QCD radiative correction to
  color-octet $J/\psi$ inclusive production at B Factories}, Phys.Rev. D81
  (2010) 034015.
\newblock \href {http://arxiv.org/abs/0911.2166} {\path{arXiv:0911.2166}},
  \href {http://dx.doi.org/10.1103/PhysRevD.81.034015}
  {\path{doi:10.1103/PhysRevD.81.034015}}.

\end{thebibliography}







\end{document}